\documentstyle[aps,eqsecnum,preprint]{revtex}
\begin{document}
\draft
\title{{\bf Spectra of regular quantum graphs}}
\author{Yu. Dabaghian, R. V. Jensen, and R. Bl\"umel}
\address{Department of Physics, Wesleyan University,\\
Middletown, CT 06459-0155, USA}
\date{\today}
\maketitle

\begin{abstract}
We consider a class of simple quasi one-dimensional classically
non-integrable systems which capture the essence of the periodic
orbit structure of general hyperbolic nonintegrable dynamical
systems.
Their behavior is simple enough to allow a detailed investigation
of both classical and quantum regimes.
Despite their classical chaoticity, these systems exhibit a
``nonintegrable analog'' of the Einstein-Brillouin-Keller
quantization formula which
provides their spectra explicitly, state by state, by
means of convergent periodic orbit expansions.
\end{abstract}

\pacs{05.45.Mt, 03.65.Sq, 02.30.Lt}

\section{introduction}
Very few quantum systems can be solved explicitly.
Among them are the standard textbook examples, such as the
harmonic oscillator or the hydrogen atom \cite{LL}.
In all of these cases the spectrum of the quantum system is
obtained as an explicit analytical formula of the form
``$E_{n}=...$'', where $n$ is the quantum number of the system.
This procedure fails already for some of the simplest quantum
systems, which are still considered elementary textbook problems.
An example is a quantum particle in a box with a step potential
inside, as shown in Fig.~1.
Even for the simple problem of Fig.~1, explicit analytical
solutions of the form ``$E_{n}=...$'' are no longer available
since the problem leads to a transcendental spectral equation.
The recommended method of solution is either numerical or graphical
\cite{LL,Messia,Flugge}.
We recently found a way \cite{Prima,Opus,Sutra} of obtaining
explicit analytical solutions of a wide class of problems such
as the one shown in Fig.~1, thus obtaining an explicit analytical
solution of textbook problems which until now were
relegated to numerical or graphical solution techniques.
Our methods are also a step forward in
the mathematical theory of almost periodic functions
\cite{Bohr} since
we obtain explicit formulae for the zeros of a wide class
of almost periodic functions. Furthermore,
the classical dynamics of the quantum systems to be
discussed in this paper is chaotic.
Since it may well be true in general
that the quantized versions of
classically chaotic systems do not admit for the existence of
quantum numbers (see, e.g.,
\cite{numbers,topent} for a detailed
discussion of this important point),
our ``$E_{n}=...$'' spectral
formulae, containing an explicit quantum number $n$, may come as
a surprise.
At this point we feel that it is important to stress
that our results are not
conjectures, approximations or merely formal identities.
Our results are exact, explicit, convergent
periodic orbit expansions that can be cast into the form of
mathematical theorems. We will publish the rigorous mathematical
underpinnings of our results elsewhere \cite{JMP}.

It is well known \cite{Gutzw}
that the periodic orbit theory leads to completely different
approaches for quantizing integrable and non-integrable dynamical
systems. For integrable systems there is a simple procedure
\cite{Gutzw,Keller} which allows us
to quantize the action
variables {\em individually} for each degree of freedom. The
situation is completely different for the chaotic case, where
the periodic orbit theory \cite{Gutzw} allows us to evaluate
only certain {\it global} characteristics of the spectrum, e.g.
the density of states
\begin{equation}
\rho (E)= \sum_{j=1}^{\infty}\delta\left(E-E_{j}\right)
\approx\bar{\rho}(E)+
\frac{1}{\pi}\mathop{\rm Im} \sum_{p}T_{p}(E)
\sum_{\nu=1}^{\infty} A_{p}^{\nu}(E)\,e^{i\nu S_{p}(E)},
\label{rho}
\end{equation}
typically only with semiclassical accuracy \cite{Sz}.
Here $\bar{\rho}(E)$ is the average density of states,
$S_{p}(E)$, $T_{p}(E)$ and $A_{p}(E)$ are correspondingly
the action, the period and the weight factor of the prime periodic
orbit labeled by $p$, and $\nu$ is the repetition index.
In this approach individual energy levels are obtained
indirectly as the singularities of the sum (\ref{rho}).
As for the idea of expressing them directly in terms of the
periodic orbits, M. V. Berry wrote in 1991 \cite{Berry}:
      {\em ``...We do not know how, or even whether, the closed
             orbit sum generates the individual $\delta$s in the
   level density for chaotic systems.
   This is a serious -- perhaps shocking --
   situation, because it means
   that we are ignorant of the mechanism of quantization"}.

In the case of quantum graphs Berry's question can be answered
definitively. The periodic orbit
sums representing the spectral density of
quantum graphs do provide the individual levels
in the form of $\delta$-spikes in (\ref{rho})
and only those
\cite{Roth,QGT1,QGT2,QGT3}.
In addition we showed recently \cite{Prima,Opus,Sutra}
that the answer to Berry's
question can be taken one step forward: not only do
periodic orbit expansions for quantum graphs produce
$\delta$-functions for the quantum states in the level
density, but for certain classes of quantum graphs
there also exist explicit convergent periodic orbit expansions for
individual energy levels.
Because they provide explicit formulas for
the energy levels of classically chaotic systems,
these periodic orbit expansions
may be considered as
``nonintegrable analogues'' of  the
Einstein-Brillouin-Keller (EBK) quantization formula
\cite{Gutzw,Keller} which applies to integrable systems.

This paper is organized as follows. In Sect.~II we briefly
review
the theory of quantum graphs and extend the theory by
defining ``dressed graphs'', i.e. quantum graphs with
arbitrary potentials on their bonds.
In Sect.~III we define
an important class of dressed
quantum graphs: regular quantum graphs.
Based on a detailed study of their spectral
properties in Sect.~III we derive
explicit analytical spectral formulae for regular
quantum graphs in Sect.~IV.
In Sect.~V we present a
variety of regular quantum graphs illustrating use
and convergence of the spectral formulae. In Sect.~VI
we summarize our results and conclude the paper.

\section{Dynamical networks}
Let us consider a particle moving on a quasi
one-dimensional network of bonds and vertices.
In the mathematical literature these networks are known
as graphs. They were and still
are the subject of intensive investigations
in all areas of science ranging from mathematics over
computer science to chemistry and physics.
An example of a simple graph with five vertices and seven
bonds is shown in Fig.~2.
The particle scatters
randomly at
every vertex $V_{i}$ along different bonds $B_{ij}$ which meet
at that vertex.
We shall assume that the graph contains a finite number of bonds
and vertices ($N_{B}$ and $N_{V}$ correspondingly).
The key assumption about the dynamics of the particle is that the
turning points of any trajectory of a particle moving on
the graph coincide with the
vertices of the graph, and hence the shape of the trajectories is
determined {\em uniquely} by the geometry of the graph.
The trajectories of the particle are simply the joint sequences
of graph bonds, which are easily described and enumerated. For
instance, every trajectory can be represented by a sequence of
$N_{B}$ symbols, each one of which corresponds to a certain bond
\cite{Nova}.
Since the trajectories correspond to various bond sequences, every
trajectory is described by a code word consisting of $N_{B}$ symbols.

We ``dress'' the bonds $B_{ij}$ of the graph with potentials,
$U_{ij}(x)$, which may affect the way a
particle moves along the bonds.
However, it is required that these dressings do
not violate the geometry of the particle trajectories, i.e. do
not add turning points, other than the original vertices of
the graph. This condition is required to hold at all energies.
To comply with this requirement
the bond potentials are
allowed to depend on the energy $E$ of the particle, i.e.
$U_{ij}=U_{ij}(x,E)$, such that
$E>U_{ij}(x,E)$ is fulfilled for all $E$ and all $i,j$.
This will in fact lead to many additional simplifications which
have a deep physical meaning in the context of the semiclassical
periodic orbit theory \cite{Prima,Opus,Sutra,Nova,PSM,Tom}.

The shapes of the trajectories, and in particular of the periodic
orbits, becomes increasingly complicated as their lengths grow.
This is what makes them similar to the generic (dynamical) chaotic
systems.
In fact, the
number of possible
periodic orbits increases exponentially
with their lengths,
(or, equivalently, the number of vertex scatterings)
with a rate which depends only on the topology of the graph.
Every graph $\Gamma$ can be characterized by its topological
entropy (global average
rate of exponential proliferation of periodic
orbits)
\begin{equation}
\Lambda_{\Gamma}=\lim_{l\rightarrow\infty}
{\ln\left[\#(l)\right]\over l},
\label{topo}
\end{equation}
where $l$ characterizes the lengths of the periodic orbits in
terms of
the lengths of their code words
and $\#(l)$ is the total number of periodic orbits of length
$\leq l$
\cite{topent}.
Since the phase space of the system is bounded, the dynamics
of the
particle is mixing \cite{QGT1}.
Hence the structure of the periodic orbit set on dynamical
networks
closely imitates the behavior of the closed trajectories of
generic
chaotic systems \cite{AA,Arnold}.
On the other hand, dynamical networks can be easily quantized
\cite{Prima,Opus,Sutra,QGT2,QGT3,Nova,Orsay}, which makes them
very convenient models for
studying various aspects of quantum chaology.

The details of the classical dynamics on graphs are discussed
in numerous publications \cite{QGT1,Gaspard}.
Below we investigate the quantum-mechanical
description of these systems. In particular, we shall discuss
their
spectra in the context of the periodic orbit theory.
So let us now briefly outline some details of the graph
quantization
procedure, which will be used in the subsequent discussion.

A quantum graph system is a {\em quantum} particle
which moves on a one-dimensional network $\Gamma$ dressed with
the potentials $U_{ij}(x,E)$.
Below we shall consider the case of {\em scaling}
potentials discussed
in \cite{Sutra,RS1,RS2,RS3},
\begin{equation}
U_{ij}(E)=\lambda_{ij}E, \ \ \lambda_{ij}=\lambda_{ji},
\label{pot}
\end{equation}
where the
$\lambda_{ij}$'s are constants. This choice of the dressing
potentials allows us to avoid certain mathematical complications,
which are irrelevant for the physical context
of our discussion. For
more details on scaling potentials and their relevance in the
semiclassical periodic orbit analysis see
\cite{Prima,Sutra,Nova}.

The Schr\"{o}dinger equation for graphs with the
potentials (\ref{pot})
can be written as
\begin{equation}
\hat \pi_{ij}^{2}\psi_{ij}(x)=\beta_{ij}^{2}E\psi_{ij}(x),
\label{schred}
\end{equation}
where
\begin{equation}
\hat \pi_{ij}=-i\frac{d}{dx}-A_{ij}
\label{mom}
\end{equation}
is the generalized momentum operator and
$\beta_{ij}^{2}=1-\lambda_{ij}$.
The coordinate $0\leq x\leq L_{ij}$ is
measured along $B_{ij}$ from $i$
to $j$, and $L_{ij}=L_{ji}$ is the length
of the bond. The magnetic
field vector potential $A_{ij}=-A_{ji}$ is
assumed to be a constant, real
matrix, and may be used as a tool for braking
the time-reversal symmetry.

Classically, the particle can travel along the
bond $B_{ij}$ if its
energy is above the scaled potential height,
$E>U_{ij}(E)$,
($\lambda_{ij}<1$).
In this case the solution of equation (\ref{schred})
on the bond
$B_{ij}$ is a combination of the free waves,
\begin{equation}
\psi_{ij}(x)=a_{ij}\frac{e^{i\left(-\beta_{ij}k+A_{ij}
\right) x}}
{\sqrt{\beta_{ij}k}}+b_{ij}\frac{e^{i\left(\beta_{ij}k+
A_{ij}\right) x}}
{\sqrt{\beta_{ij}k}},
\label{psi}
\end{equation}
where $k=\sqrt{E}$ and
the factors $(\beta_{ij}k)^{-1/2}$ are introduced to separate the
physically meaningful flux amplitudes from the
coefficients $a_{ij}$ and $b_{ij}$.
In the opposite case of $\lambda_{ij}>1$, the bond $B_{ij}$
carries a
linear combination of tunneling solutions. Due
to the scaling assumption, there is no transition between these two
cases as a function of $E$. From now on we shall assume
that the energy $E$
is kept above the maximal scaled potential height,
\begin{equation}
\lambda_{ij}<1,\ \ i,j=1,...,N_{V}.
\label{above}
\end{equation}
At every vertex $V_{i}$, the bond wave functions satisfy the boundary
conditions,
\begin{eqnarray}
\psi_{ij}(x &=&0)=\varphi_{i}C_{ij},\
\label{bound} \cr
\sum_{j=1}^{N_{V}}C_{ij}\hat{\pi}_{ij}\psi_{ij}(x) &\mid &_{x=0}=
-i\lambda_{i}\varphi_{i},
\label{bound1}
\end{eqnarray}
for all $i,j=1,...,N_{V}$.
Here $C_{ij}$ is the connectivity matrix
of the graph, $\varphi_{i}$ is the
value of the wave function at the
vertex $V_{i}$, and the $\lambda_{i}$'s
are free parameters of the problem,
scaled as $\lambda_{i}=\lambda_{i}^{0}k$
(see Appendix). Note that the double-indexed
scaling constants $\lambda_{ij}$ refer to the bonds,
whereas the single-indexed constants
$\lambda_i$ refer to the scattering strengths at
the vertices. We believe that this notation is
natural and does not lead to confusion.

The conditions (\ref{bound})
are consistent only for a discrete set
of energy levels, $E_n=k_n^2$
which define the spectrum of the dressed
quantum graph problem (\ref{schred}) and (\ref{bound}).
As shown in
\cite{Sutra,QGT1,QGT2,QGT3,Orsay} (see Appendix),
using the scattering quantization approach
\cite{Uzi}, one obtains the
spectral equation for any quantum graph problem in the form
\begin{equation}
\Delta (k)=\det[1-S(k)]=0,
\label{det}
\end{equation}
where $S(k)$ is the finite unitary
graph scattering matrix \cite{QGT1}.
The indices that define the matrix
elements $S_{IJ}$ of the matrix $S$
correspond to the graph bonds. It is important that the bond
$B_{I}\equiv B_{ij}$ is considered
to be different from the (geometrically
identical) reversed bond $B_{I'}\equiv B_{ji}$,
so the bonds of the graph
are directed \cite{Prima,Opus,Sutra,QGT1,QGT3}.
Hence the dimensionality of the scattering matrix
is $2N_{B}\times 2N_{B}$.
It is shown in the Appendix that $S=TD(k)$, where $T$
is a constant
$2N_{B}\times 2N_{B}$ unitary matrix and $D$ is a
diagonal unitary matrix,
whose matrix elements are given by
\begin{equation}
D_{IJ}=\delta_{IJ}e^{i\left(\beta_{I}k+A_{I}\right)L_{I}},
\ \ I=1,...,2N_{B}.
\label{d}
\end{equation}
Since $\Delta(k)$ is a complex function, it is convenient to define
the spectrum via the zeros of its absolute value,
\begin{equation}
|\Delta(k)|=e^{-i\Theta_{0}(k)}\Delta(k),
\label{modul}
\end{equation}
where $\Theta_{0}(k)$ is the complex phase of $\Delta(k)$.
The logarithmic derivative of $|\Delta(k)|$
produces a delta-peak for every
one of its roots,
\begin{equation}
-\frac{1}{\pi}\mathop{\rm Im}\lim_{\epsilon
\rightarrow 0}\frac{d}{dk}
\ln |\det\left[1-S\left(k+i\epsilon\right)\right]| =
\sum_{n=1}^{\infty}\delta(k-k_{n}),
\label{delta}
\end{equation}
which, by definition, is the density of the
momentum states $\rho (k)$
\cite{Sutra}.
On the other hand, using (\ref{modul}) and
expanding the logarithm of
the determinant (\ref{det}), the density of
states can be written as
\begin{equation}
\rho (k)=\frac{1}{\pi}\frac{d\Theta_{0}(k)}{dk}+
\frac{1}{\pi}\mathop{\rm Im}
\frac{d}{dk}\sum_{n=1}^{\infty}
\frac{1}{n}\mathop{\rm Tr}\left[S(k)\right]^{n}.
\label{exp}
\end{equation}
Then it can be easily seen from
the structure of the scattering matrix $S$
\cite{QGT1,Orsay}, that the matrix
elements of its {\em n}-th power are
defined on connected sequences of $n$
bonds and the {\em trace}
of $S^{n}$ generates terms defined on
{\em closed} connected sequences
of {\em n} bonds \cite{Sutra,QGT2,Orsay,Gaspard}.

These periodic connected sequences of $n$ bonds $B_{ij}$,
can be viewed
as the periodic orbits traced by a classical point particle
moving on
the graph.
Note that the phase of the exponent in (\ref{d}) is exactly
the action
of a classical point particle's trajectory traversing the
bond $B_{I}$,
\begin{equation}
{\cal{S}_{I}}=\int_{B_{I}}\left(\beta_{I}k+A_{I}\right)dx=
\left(\beta_{I}k+A_{I}\right)L_{I}.
\label{action}
\end{equation}
Thus the semiclassical transition amplitudes
$e^{i{\cal{S}}_{I}}$
between the vertices connected by the bond $B_{I}$
determine the
scattering matrix $S(k)$.
As a consequence
\cite{Prima,Opus,Sutra,QGT1,Nova}, the ``closed bond
sequence expansion''
(\ref{exp}) can be written explicitly as a periodic
orbit expansion in terms of the phases (\ref{action}),
\begin{equation}
\rho (k)=\bar{\rho}(k)+\frac{1}{\pi}\mathop{\rm Re}
\sum_{p}S_{p}^{0}
\sum_{\nu=1}^{\infty}A_{p}^{\nu}\,e^{i\nu S_{p}^{0}k},
\label{ro}
\end{equation}
where $S_{p}^{0}$ is the $k$-independent ``action length'' of
the orbit $p$,
\begin{equation}
S_{p}=\sum_{p}\beta_{ij}L_{ij}k\equiv S_{p}^{0}\,k,
\end{equation}
and $A_{p}$ is its weight containing the constant factor
$e^{i\sum_{p}A_{ij}L_{ij}}$.
Due to the scaling assumption (see Appendix) the
weight factor $A_{p}$ is $k$-independent.
The first term in this expression corresponds to the average
density of states of momentum $\bar{\rho}(k)$,
\begin{equation}
\bar \rho(k)=\frac{1}{\pi}\frac{d\Theta_{0}(k)}{dk},
\label{roave}
\end{equation}
while the periodic orbit sum in (\ref{ro}) describes the
fluctuations around the average.

The periodic orbit expansion for the staircase function
\begin{equation}
N(k)=\sum_{n=1}^{\infty}\Theta (k-k_{n})
\label{N}
\end{equation}
can be obtained by direct
integration of (\ref{delta}) and (\ref{ro}).
We obtain
\begin{equation}
N(k)=\bar{N}(k)+ \tilde N(k),
\label{sta}
\end{equation}
where the first term,
\begin{equation}
\bar{N}(k)=\int_{0}^{k}\bar\rho(k')\,dk'+\bar N(0)
\label{avestair}
\end{equation}
represents the average behavior of the staircase and
\begin{equation}
\tilde N(k)=\mathop{\rm Im}\frac{1}{\pi}
\sum_{p}\sum_{\nu =1}^{\infty}
\frac{A_{p}^{\nu}}{\nu}e^{i\nu S_{p}^{0}k}
\label{var}
\end{equation}
describes zero-mean oscillations around the average.

As discussed in the Introduction (see also
\cite{Prima,Opus,Sutra,QGT1,Gaspard}),
quantum graphs are chaotic in the classical limit. The classical
scattering probabilities are obtained in the limit
$\hbar\rightarrow 0$ from the quantum mechanical
transition amplitudes
\cite{Prima,Opus,Sutra}, (see Appendix).
In the scaling case they are $k$-independent and
thus the
quantum scattering amplitudes do not depend on
$\hbar$ at all.
They determine simultaneously the quantum and
the classical scattering
probabilities.

\section{Regular graphs and their spectra}
The spectral determinant is a polynomial of degree $2N_{B}$
of the matrix elements of $S$.
It was shown in \cite{Sutra} that the total phase
of this polynomial
is
\begin{equation}
\Theta_{0} (k)=\frac{1}{2}
\mathop{\rm Im}\ln\det S(k)=kS_{0}-\pi\gamma_{0},
\label{theta}
\end{equation}
where $S_{0}=\sum_{(ij)}L_{ij}\beta_{ij}$ is the total action
length of the graph $\Gamma $ and
\begin{equation}
\gamma_{0}=(N_{B}+N_{V})/2+\frac{1}{\pi}\sum_{i=1}^{N_{V}}
\arctan\left(\frac{\lambda^{0}_{i}}{v_{i}}\right),
\label{gamma}
\end{equation}
where
\begin{equation}
v_i=\sum_j \, C_{ij}\, \beta_{ij}.
\end{equation}
Therefore the average density of states is a constant,
\begin{equation}
\bar \rho=\frac{1}{\pi}\frac{d}{dk}\Theta_{0} (k)=
\frac{S_{0}}{\pi},
\label{roa}
\end{equation}
and the average staircase (\ref{avestair}) is
\begin{equation}
\bar N(k)=\frac{S_{0}}{\pi}k+\bar N(0).
\label{Nave}
\end{equation}
The spectral equation $|\Delta(k)|=0$ can be written
in the form
\begin{equation}
\cos \left(S_{0}k-\pi \gamma_{0} \right)=
\sum_{i=1}^{N_{\Gamma}}a_{i}\cos(S_{i}k-\pi\gamma_{i}),
\label{eqn}
\end{equation}
where the frequencies $S_{i}< S_{0}$ are combinations
of the reduced classical actions $S^{0}_{ij}=\beta_{ij}L_{ij}$,
and $\gamma_{0}$, $\gamma_{i}$ are constants.
The number $N_{\Gamma}$ of terms in (\ref{eqn}) is bounded by
$N_{\Gamma}\leq 3^{N_{B}}$ \cite{Sutra}.

The frequency $S_{0}$ in $\Theta_{0}(k)$ is the
{\em largest} frequency in the expansion (\ref{eqn}).
While it is the only characteristic of the graph contained in
the left-hand side of (\ref{eqn}), the right-hand side
\begin{equation}
\Phi (k) \equiv \sum_{i=1}^{N_{\Gamma}}a_{i}\cos(S_{i}k-
\pi\gamma_{i}),
\label{fi}
\end{equation}
contains the complete information about the graph system.
We call $\Phi (k)$ the {\em characteristic function} of
the graph.

A graph $\Gamma$ is called {\em regular}
\cite{Prima,Opus,Sutra},
if its characteristic function $\Phi(k)$ satisfies
\begin{equation}
\sum_{i=1}^{N_{\Gamma}}|a_{i}|\equiv\alpha<1.
\label{reg}
\end{equation}
 For regular graphs the spectral equation (\ref{eqn}) can
be solved formally \cite{Prima,Opus,Sutra} to yield the
implicit equation of its eigenvalues,
\begin{equation}
k_{n}={\frac{\pi}{S_0}}\left[n+\mu+\gamma_{0}\right] +
{\frac{1}{S_0}}
\cases{
\arccos[\Phi(k_n)], &for $n+\mu$ even, \cr
\pi-\arccos[\Phi(k_n)], &for $n+\mu$ odd, \cr}
\label{levels}
\end{equation}
where $\mu$ is a fixed integer, chosen such that $k_1$ is the
first positive solution of (\ref{eqn}).
The index $n\in{\bf N}$ labels the roots of (\ref{eqn}) in their
natural sequence.

 From the implicit form (\ref{levels}) it follows immediately that
since the second term in (\ref{levels}) is bounded by $\pi/S_{0}$,
the deviations of solutions to this equation from the points
\begin{equation}
\hat k_n =\frac{\pi}{S_0}(n+\mu+\gamma_0+1)
\label{mid}
\end{equation}
never exceeds $\pi /S_{0}$ in absolute value for any $n$.
Below it will turn out that the quantities $\hat k_n$ are very
important since they determine the root structure of
(\ref{eqn}).

The roots $k_{n}$ can be decomposed into an average part $\bar k_{n}$
and a fluctuating part $\tilde k_{n}$. From (\ref{levels}) we obtain
\begin{equation}
k_{n}=\bar k_{n}+\tilde k_{n},
\label{average}
\end{equation}
where
\begin{equation}
\bar k_{n}={\frac{\pi}{S_0}}\left[n+\mu+\gamma_{0}+1/2\right],
\label{avek}
\end{equation}
and
\begin{equation}
\tilde k_{n}={\frac{(-1)^{n+\mu}}{S_0}}
\left\{\arccos[\Phi(k_n)]-\frac{\pi}{2}\right\}.
\label{fluct}
\end{equation}
Note that the
constant $\mu+\gamma_{0}$ can be related to the initial value
$\bar N(0)$ of the average staircase function (\ref{Nave}).
Consider the integral
\begin{equation}
\lim_{n\rightarrow\infty}\frac{1}{\hat k_{n}}
\int_{0}^{\hat k_{n}}N(k')dk'.
\label{integral}
\end{equation}
The integration in (\ref{integral}) can be easily performed
due to the simple form (\ref{N}) of the function $N(k)$:
\begin{eqnarray}
\frac{1}{\hat k_{n}}\int_{0}^{\hat k_{n}}N(k') dk'=
n-\frac{1}{\hat k_{n}}\sum_{i=1}^{n}k_{i},
\label{integr}
\end{eqnarray}
since there are $n$ roots to the left of $\hat k_{n}$.
The fluctuations of both $N(k)$ and $k_{n}$ around their averages
have zero mean, so in the limit of $n\gg 1$ one can use $\bar N(k)$
and $\bar k_{n}$ instead of $N(k)$ and $k_{n}$ in (\ref{integr}),
and write
\begin{equation}
\frac{1}{\hat k_{n}}\int_{0}^{\hat k_{n}}\bar N(k')dk'=
n-\frac{1}{\hat k_{n}}\sum_{i=1}^{n}\bar k_{i}, \ \ n\gg 1.
\label{a10}
\end{equation}
Using the explicit forms of $\hat k_{n}$,
$\bar k_{n}$ and $\bar N(k)$,
we obtain
\begin{equation}
\bar{N}(0)+\frac{1}{2}(n+\mu+\gamma_{0}+1)=
\frac{n}{n+\mu+\gamma_{0}+1}
\left(\frac{n}{2}+\mu+\gamma_{0}+1\right).
\end{equation}
Expanding the right-hand side and keeping
terms up to order $1/n$
yields
\begin{equation}
\bar{N}(0)+\frac{1}{2}(n+\mu+\gamma_{0}+1)=n-
\left(1-\frac{\mu+\gamma_{0}+1}{n}\right)
\left(\frac{n}{2}+\mu+\gamma_{0}+1\right).
\label{expan}
\end{equation}
The terms proportional to $n$ cancel out.
Comparing the constants in (\ref{expan}) yields
\begin{equation}
\bar{N}(0)=-(\mu+\gamma_{0}+1).
\label{n0}
\end{equation}
One can verify by direct substitution that
\begin{eqnarray}
\bar N(\hat k_{n})=n,
\label{inverse}
\end{eqnarray}
which implies that the function (\ref{mid})
is the inverse of the average
staircase (\ref{Nave}).
One can also view the points $\hat k_{n}$ as the intersection
points of the staircase (\ref{N}) and its average (\ref{Nave}),
\begin{eqnarray}
\bar N(\hat k_{n})=N(\hat k_{n})=n,
\label{cross}
\end{eqnarray}
so at the points $\hat k_{n}$ the fluctuations
$\tilde N(k)$ of the
spectral staircase vanish,
\begin{equation}
\tilde N(\hat k_{n})=\mathop{\rm Im}\frac{1}{\pi}
\sum_{p}\sum_{\nu =1}^{\infty}
\frac{A_{p}^{\nu}}{\nu}e^{i\nu S_{p}^{0}\hat k_{n}}
=0.
\label{vanish}
\end{equation}
Geometrically, (\ref{vanish}) means that the average staircase
$\bar N(k)$ intersects every step of the staircase function
$N(k)$. Hence we call $\bar N(k)$ the {\em piercing average}.
This is illustrated in Fig.~3 which shows the spectral
staircase $N(k)$ for the scaling step potential
shown in Fig.~1 and discussed in more detail in Sect.~V,
Example 1, below. We used the parameters $\lambda=1/2$ and
$b=0.3$. Also shown is the average staircase $\bar N(k)$
for this case. It clearly pierces all the steps of
$N(k)$ providing an example of a system with a piercing
average.

Since $\Phi(k)$ contains only frequencies smaller than $S_{0}$,
every {\it open} interval $I_{n}=(\hat k_{n-1},\hat k_{n})$
contains only one root of (\ref{eqn}), namely $k_{n}$. Thus
the $\hat k_n$ play the role of separating points between
adjacent roots.
\cite{Prima,Opus,Sutra,JMP}.
Moreover, because of (\ref{reg}), the ``allowed zones''
$R_{n}\subset I_{n}$, where the roots $k_{n}$ can be found,
narrow to
\begin{equation}
k_{n}\in R_{n}\equiv\left(\frac{\pi}{S_{0}}
\left(n+\mu+\gamma_{0}+u\right),
\frac{\pi}{S_{0}}\left(n+\mu+\gamma_{0}+1-u\right)\right),
\label{int}
\end{equation}
where $u=\arccos(\alpha)/S_{0}$. Correspondingly,
there are forbidden
regions $F_{n}$,
\begin{equation}
     F_{n}\equiv\left(\frac{\pi}{S_{0}}
\left(n+\mu+\gamma_{0}-u\right),
\frac{\pi}{S_{0}}\left(n+\mu+\gamma_{0}+1+u\right)\right),
\label{zap}
\end{equation}
where roots of (\ref{eqn}) never appear. In the limit
$\alpha\rightarrow 1$ ($u\rightarrow 0$) the allowed zones $R_n$
tend to occupy
the whole root interval, $R_n\rightarrow I_n$.

\section{Spectral Formulae}
Once the existence of separating points $\hat k_n$
has been established, it is
possible to obtain an exact periodic orbit expansion
{\em separately
for every root of} (\ref{det}). The derivation is
based on the identity
\begin{equation}
k_{n}=\int_{\hat{k}_{n-1}}^{\hat{k}_{n}}k\rho (k)dk.
\label{root}
\end{equation}
Substituting the exact periodic orbit expansion (\ref{ro})
for $\rho(k)$ into
(\ref{root}), yields
\begin{eqnarray}
k_{n}=\int_{\hat{k}_{n-1}}^{\hat{k}_{n}}k
\frac{S_{0}}{\pi}dk+\frac{1}{\pi}
\int_{\hat{k}_{n-1}}^{\hat{k}_{n}}k
\mathop{\rm Re}\sum_{p}S_{p}^{0}
\sum_{\nu =1}^{\infty}A_{p}^{\nu}\,e^{i\nu
S_{p}^{0}k}dk=\frac{\pi}{S_{0}}(n+\mu+\gamma_{0}+1/2)
\cr+
\hat{k}_{n}\mathop{\rm Im}\frac{1}{\pi}
\sum_{p}\sum_{\nu=1}^{\infty}
\frac{A_{p}^{\nu}}{\nu}e^{i\nu S_{p}^{0}
\hat{k}_{n}}-\hat{k}_{n-1}
\mathop{\rm Im}\frac{1}{\pi}\sum_{p}
\sum_{\nu=1}^{\infty}\frac{A_{p}^{\nu}}
{\nu}e^{i\nu S_{p}^{0}\hat{k}_{n-1}}
\cr+
\mathop{\rm Re}\frac{1}{\pi}\sum_{p}
\frac{1}{S_{p}^{0}}\sum_{\nu=1}^{\infty}
\frac{A_{p}^{\nu}}{\nu ^{2}}\left(e^{i\nu S_{p}^{0}\hat{k}_{n}}-
e^{i\nu S_{p}^{0}\hat{k}_{n-1}}\right).
\label{first}
\end{eqnarray}
Using (\ref{vanish}) we simplify (\ref{first}) to obtain
\begin{eqnarray}
k_{n}=\frac{\pi}{S_{0}}(n+\mu+\gamma_{0}+1/2)
-\frac{1}{\pi}\mathop{\rm Im}\sum_{p}
\frac{2}{S_{p}^{0}}\sum_{\nu =1}^{\infty}
\frac{A_{p}^{\nu}}{\nu ^{2}}\sin
\left[{\nu \omega_p\over 2}\right]
e^{i\nu \omega_p \left(n+\mu+\gamma_{0}+1/2\right)},
\label{kn}
\end{eqnarray}
where $\omega_{p}=\pi S_{p}^{0}/S_{0}$.
The series expansion (\ref{kn}) for $k_{n}$ is more than a
formal identity.
It is rigorously convergent, however
it converges only {\em conditionally},
which means that the result of the summation depends
essentially on how the summation was performed.
Indeed, according to Riemann's well-known reordering
theorem, one can obtain
{\em any} result by rearranging the
terms of a conditionally convergent
series \cite{Hilbert}.
Hence for proper convergence of (\ref{kn}) to the exact roots of the
spectral equation (\ref{eqn}), we have to specify how the terms in
(\ref{kn}) are to be summed.

The mathematical details of the convergence properties of (\ref{kn})
are presented in \cite{JMP}. Here we mention the main result,
which states that the terms in (\ref{kn}) have to be summed
according to the length of the symbolic codes \cite{Sutra,Nova} of
the periodic orbits, and not according to their action lengths.
If (\ref{kn}) is summed in this way it not only converges, but also
converges to the exact roots $k_{n}$
of the spectral equation (\ref{eqn}).

Hence the formula (\ref{kn}) provides
an explicit representation of
the roots of the spectral equation (\ref{det}) in terms of the
geometric characteristics of the graph.
According to (\ref{avek}), the first term in (\ref{kn}) is the average
$\bar k_{n}$, and the following periodic orbit sum is an explicit
expression for the fluctuation of the root $\tilde k_{n}$.
This method is not limited to obtaining explicit analytical periodic
orbit expansions for $k_{n}$.
In fact, by using the identity
\begin{equation}
f(k_{n})=\int_{\hat{k}_{n-1}}^{\hat{k}_{n}}f(k)\rho (k)dk,
\label{general}
\end{equation}
we can obtain periodic orbit
expansions for any function of the eigenvalues
$f\left(k_{n}\right)$, for instance for the energy $E=k^{2}$.

In the simplest case where
$\lambda^{0}_{i}=0$, $A_{ij}=0$ and
$\mathop{\rm Im} A_{p}=0$ we have
\begin{eqnarray}
k_{n}=\frac{\pi}{S_{0}}n-\frac{2}{\pi}\sum_{p} \frac{1}{S_{p}^{0}}
\sum_{\nu=1}^{\infty}\frac{A_{p}^{\nu}}{\nu^{2}} \sin(\frac{1}{2}
\nu\omega_{p})\,\sin(\nu\omega_{p}n).
\label{kn0}
\end{eqnarray}
Note that in this case $k_{-n}=-k_{n}$.

Both EBK theory as well as formula (\ref{kn}) allow us to
compute energy eigenvalues explicitly. In this sense
formula (\ref{kn}) may be regarded as an analog
of the EBK quantization formula \cite{Gutzw,Keller}
for a chaotic system.
The complexity of this expansion, structurally similar to (\ref{rho}),
reflects the geometrical complexity of the periodic orbit set for
graph systems.

  Finally, for explicit calculations (see the following
Sect.~V)
it remains to
determine the
explicit form of the expansion coefficients $A_{p}$. For
some simple graphs this was done
in \cite{QGT1,Nova}. In the Appendix we solve the problem
for general dressed graphs. We
show that every passage
of an orbit $p$ from a bond $B_{ij}$ to $B_{ij'}$
through a vertex $V_{i}$
contributes a factor $\sigma_{ji,ij'}$
(a matrix element of the matrix $T$,
see Appendix) to the weight $A_{p}$ of the orbit,
\begin{eqnarray}
A_{p}=\prod\,\sigma_{jj'}^{(i)},
\label{weight}
\end{eqnarray}
where
the product is taken over the sequence
of the bonds traced by the orbit
$p$.

\section{Examples}
In (\ref{reg}) we provided a definition of
regular quantum graphs and discussed analytical
properties of their spectra in (\ref{levels})--(\ref{zap}).
The discussion of regular quantum graphs culminated
in Sect. IV with the
derivation of explicit spectral formulae for individual quantum
states of regular quantum graphs.
It is one thing to define regular quantum graphs, and it is quite
another to show that regular quantum graphs actually exist.
Examples 1--3, discussed below,
provide specific examples of
quantum graphs that are regular for all
choices of their parameters. Examples 4 and 5 present quantum graphs
that exhibit both regular and irregular regimes. Finally,
examples 6 and 7 provide illustrations of
a new class of quantum graphs, {\em marginal} quantum graphs,
for which $\sum_{i=1}^{N_{\Gamma}}|a_{i}|=1$. Except for special
choices of their dressing potentials these graphs can
still be accommodated
within the mathematical framework set up in Sections III and IV, and
also admit an explicit representation of their spectra according
to the spectral formulae derived in Sect. IV.

{\bf Example 1}. Scaling step potential inside of a box.
Let us consider the case of a particle confined to a box
$0<x<1$ containing the
scaling step potential
(see Fig.~1 and Fig.~4a),
\begin{equation}
U(x) =\cases{0, &for $0<x\leq b$, \cr
     \lambda_{23}E, &for $b<x<1$.\cr}
\label{pot3}
\end{equation}
This is equivalent to a three-vertex linear chain graph (Fig.~4a'),
with $\lambda_{2}=0$, $A_{ij}=0$
and Dirichlet boundary conditions at $V_{1}$
and $V_{3}$. This example is also
discussed in \cite{Prima,Opus,Sutra,Nova,JMPpub}.
In this case
the spectral equation (\ref{eqn}) can be written as
\cite{Prima,Opus}
\begin{equation}
        \sin\left[k(S^{0}_{21}+S^{0}_{23})\right]=
        r\sin\left[k(S^{0}_{21}-S^{0}_{23})\right],
\label{old}
\end{equation}
where
\begin{equation}
r=\frac{1-\beta_{23}}{1+\beta_{23}}<1
\label{r23}
\end{equation}
is the reflection coefficient at the vertex $V_{2}$.
Hence the regularity condition (\ref{reg})
is automatically satisfied
and this graph is always regular.
In Sect. IV we already discussed the convergence
properties of (\ref{kn}),
including the fact that a rigorous mathematical
proof for the convergence
of (\ref{kn}) exists \cite{JMP}.
Here we present solid numerical evidence for
the convergence of
(\ref{kn}) in the context of the scaling step-potential
(\ref{pot3}).
As discussed in \cite{Prima,Opus,Sutra,Nova},
every periodic orbit in the
potential (\ref{pot3}) can be
described by a binary code word. Figure~5
shows the relative error
$\epsilon_n^{(l)}=|k_n^{(l)}-k_n|/k_n$,
$n=1,10,100$,
of the result $k_n^{(l)}$ predicted
by (\ref{kn}) compared with the numerically
obtained exact result $k_n$ as a
function of the binary code length $l$ of
the orbits used in the expansion (\ref{kn}).
We used $b=0.3$, $\lambda=1/2$. Figure~5 also
demonstrates that using all
periodic orbits up to binary
code length $l\sim 150$, we obtain an accuracy
on the order of
$10^{-4}-10^{-7}$ for the roots $k_n$ of
(\ref{old}).
Although the convergence of the series
is slow (according to Fig.~5 it is approximately
on the order of
$\sim 1/l^{2}$ on average), one can obtain a
fairly good estimate for the roots
using all orbits of code length 20 and smaller.

{\bf Example 2}.
Scaling $\delta$-function in a box.
This potential, shown in Fig.~4b, is
again equivalent to a three-vertex quantum graph.
This time, however, the potentials
on the bonds are identically zero,
whereas the vertex $V_{2}$ is dressed
with a scaling $\delta$-function
of strength $\lambda_2=\lambda_2^0 k> 0$.
At the open ends we apply Dirichlet boundary conditions.
In this case the
spectral equation (\ref{eqn}) is
\begin{equation}
    \cos \left[k(S^{0}_{21}+S^{0}_{23})-\pi\gamma_{0}\right]=
    -|r|\cos\left[k(S^{0}_{21}-S^{0}_{23})\right],
\label{3hyd}
\end{equation}
where
\begin{eqnarray}
\gamma_0=1-{1\over\pi}\arcsin\left(
{2\over\sqrt{4+(\lambda_2^0)^2}}\right)
\label{phase}
\end{eqnarray}
and the reflection coefficient $r$ is given by
\begin{equation}
r={\lambda_2^0\over 2i-\lambda_2^0}.
\label{refl}
\end{equation}
Because of $|r|<1$,
the characteristic function of (\ref{3hyd}) also satisfies the
regularity condition (\ref{reg}). Therefore the scaling $\delta$
function in a box is another example of a regular quantum graph.

{\bf Example 3}.
Combined scaling step and scaling
$\delta$-potential in a box (Fig.~4c).
This is equivalent to a three-vertex
dressed linear graph (Fig.~4c'),
with $\lambda_2=\lambda_2^0 k> 0$.
The spectral equation (\ref{eqn})
is
\begin{equation}
     \cos \left[k(S^{0}_{21}+S^{0}_{23})-\pi\gamma_{0}\right]=
     a_1\cos \left[k(S^{0}_{21}-S^{0}_{23})-\pi\gamma_{1}\right],
\label{3hydra}
\end{equation}
where
\begin{eqnarray}
\gamma_{0}=1-{1\over \pi}\arcsin\left({\beta_{12}+\beta_{23}\over
\sqrt{(\beta_{12}+\beta_{23})^2+(\lambda_2^0)^2}}\right)
\cr
\gamma_{1}=1-{1\over \pi}\arcsin\left({\beta_{12}-\beta_{23}\over
\sqrt{(\beta_{12}-\beta_{23})^2+(\lambda_2^0)^2}}\right)
\label{phases}
\end{eqnarray}
and the coefficient $a_1$ is
\begin{equation}
a_1=\sqrt{{(\beta_{21}-\beta_{23})^2+(\lambda_2^0)^2\over
(\beta_{21}+\beta_{23})^2+(\lambda_2^0)^2}}< 1.
\label{ref1}
\end{equation}
So the characteristic function of (\ref{3hydra}) once again satisfies
the regularity condition for any linear three-vertex graph with
nontrivial bond potentials ($\beta_{21}^{2}+\beta_{23}^{2}\neq 0$)
\cite{Sutra}.

Quantum graphs which are regular
for all of their parameter values are
quite exceptional. In general
quantum graphs may have a regular regime
for a certain range of the parameter values
or the regular regime may not exist
at all. The following Example 4 illustrates this point.

{\bf Example 4}. Two scaling steps in a box (Fig.~4d).
As an example of a graph which has both a regular and an
irregular regime, let us consider a quantum particle in
a box with two scaling steps (Fig.~4d), equivalent to the
four-vertex linear graph shown in Fig.~4d'. Since there are
no $\delta$-functions present we have
$\lambda_{2}=\lambda_{3}=0$. We assume
Dirichlet boundary conditions at the dead ends of this graph.
In this case
the spectral equation (\ref{eqn}) is given by
\begin{eqnarray}
\sin(S_{0} k)=-r_{2}\sin(kS_{1})-
r_{2}r_{3}\sin(kS_{2})+r_{3} \sin(kS_{3}),
\label{4vertex}
\end{eqnarray}
where
\begin{eqnarray}
S_0&=S_{21}^0+S_{23}^0+S_{34}^0,\ \ \
S_{1}=S_{23}^0+S_{34}^0-S_{21}^0, \cr
S_{2}&=S_{21}^0+S_{34}^0-S_{23}^0,\ \ \
S_{3}=S_{21}^0+S_{23}^0-S_{34}^0
\end{eqnarray}
and $r_2$, $r_3$ are the reflection coefficients
\begin{equation}
r_2={\beta_{12}-\beta_{23}\over \beta_{12}+\beta_{23}},\ \ \
r_3={\beta_{23}-\beta_{34}\over \beta_{23}+\beta_{34}}
\end{equation}
at the corresponding vertices $V_i$. For
\begin{equation}
|r_{3}|+|r_{2}r_{3}|+|r_{2}|<1,
\label{max0}
\end{equation}
the four-vertex linear graph (Fig.~4d', Fig.~6a) is regular.
The regularity condition (\ref{max0}) is fulfilled
in a diamond-shaped region of
($r_2,r_3$) parameter space shown as the shaded area
in Fig.~6b.
The difference between the regular
and the irregular regimes is clearly reflected
in the staircase functions. Figure~7a
shows the
staircase function $N(k)$ together with the average
staircase $\bar N(k)$ in the regular regime
for the parameter combination
$r_{1}=0.2$ and $r_{3}=0.3$.
The piercing-average condition is clearly satisfied. Figure~7b
shows the
staircase function $N(k)$ together with the average
staircase $\bar N(k)$ in the irregular regime
for the parameter combination
$r_{1}=0.98$ and $r_{3}=0.99$.
In this case the piercing-average condition is clearly
violated, consistent with the irregular nature of this
regime.

{\bf Example 5}. Two scaling $\delta$-functions in a box
(Fig.~4e). This potential is
equivalent with the four-vertex graph shown in Fig.~4e'
with $\lambda_2=\lambda_2^0 k> 0$,
$\lambda_3=\lambda_3^0 k> 0$,
and Dirichlet boundary conditions at the dead ends
$V_1$ and $V_4$.
In this case the
spectral equation (\ref{eqn}) is given by
\begin{eqnarray}
\cos(kS_0-\pi\gamma_{0})=a_1\cos(kS_1-\pi\gamma_1)+
a_2\cos(kS_2-\pi\gamma_2)+a_3\cos(kS_3-\pi\gamma_3),
\label{4vert}
\end{eqnarray}
where
\begin{equation}
a_1={\lambda_2^0\over\sqrt{4+(\lambda_2^0)^2}},\ \ \
a_2={\lambda_3^0\over\sqrt{4+(\lambda_3^0)^2}},\ \ \
a_3={\lambda_2^0\lambda_3^0\over\sqrt{[4+(\lambda_2^0)^2]
     [4+(\lambda_3^0)^2]}}
\label{ref3}
\end{equation}
and
\begin{eqnarray}
\gamma_0&={1\over\pi}\,\arcsin\left({\lambda_2^0\lambda_3^0-4\over
      \sqrt{[4+(\lambda_2^0)^2][4+(\lambda_3^0)^2]}}\right),\cr
\gamma_1&={1\over\pi}\arcsin(a_2),\cr
\gamma_2&={1\over\pi}\arcsin(a_1),\cr
\gamma_3&={1\over 2}.
\end{eqnarray}
The sum of the amplitudes in (\ref{ref3}) ranges between
0 and 3. Therefore this system
has regular and irregular regimes.
The regular regime corresponds to a finite area
in the $(\lambda_2^0,\lambda_3^0)$ parameter space.
All linear chain graphs with a finite number of vertices
and Dirichlet boundary conditions at the two dead-end
vertices at the beginning and at the end of the graph
have a finite-measure regular regime and an irregular regime.
This fact is proved in \cite{JMP}.

A new type of graphs are {\it marginal quantum graphs}.
A marginal quantum graph is defined by
\begin{equation}
\sum_{i=1}^{N_{\Gamma}}|a_{i}|=1.
\label{regmarg}
\end{equation}
  For marginal quantum graphs,
apart from a small set of ``special'' graphs,
explicit spectral formulae
still exist. Explicit examples are
provided by
circular graphs (see Example 6)
and star graphs (see Example 7).

{\bf Example 6}. Scaling step potential in a box
with periodic boundary
conditions.
This system is identical with the
two-vertex circular graph shown in Fig.~8.
In the case of a circular graph a minor notational
problem arises since starting from vertex $V_1$, say,
vertex $V_2$ may be reached directly via two different
bonds. For the purposes of this example we solve the
problem in the following way. First we introduce a
positive sense of rotation, i.e.
mathematically positive or
counterclockwise, for the
circular graph in Fig.~8. Then we introduce the
scaling coefficient $\beta_{12}$ referring the the
bond that connects vertex $V_1$ with vertex $V_2$
traversing the graph in the mathematically positive sense.
We introduce the scaling coefficient $\beta_{21}$
which refers to the bond that connects vertex $V_2$
with vertex $V_1$, again in the mathematically positive sense.
We use the same notation for the two reduced
actions $S_{12}^0$ and
$S_{21}^0$ referring to the two
different bonds (in the mathematically
positive sense), respectively. This notation is not
confusing here, since no magnetic field is switched on
($A_{ij}=0$).
With this notation
the spectral equation is
given by
\begin{equation}
\cos(kS_0) = a_1 + a_2\, \cos(kS_2),
\label{2circ}
\end{equation}
where $S_0=S_{12}^0+S_{21}^0$,
$S_1=S_{12}^0+S_{21}^0$,
\begin{equation}
a_1={4\beta_{12}\beta_{21}\over (\beta_{12}+\beta_{21})^2}
\end{equation}
and
\begin{equation}
a_2=\left({\beta_{12}-\beta_{21}\over \beta_{12}+\beta_{21}}
\right)^2.
\end{equation}
Note that $a_1+a_2=1$. Condition (\ref{regmarg}) is
satisfied and the circular quantum graph with a
scaling step potential is marginal.

Although the strict inequality (\ref{reg}) is violated,
it is important to note that even in the marginal case
the separating points
$\hat k_{n}$, generically, are still not
solutions to (\ref{2circ}).
This happens only for special parameter combinations,
and therefore for special quantum graphs for which
the equation
\begin{equation}
(-1)^{n+\mu+1}=a_1+a_2\cos(\hat k_n S_2)
\label{2circl}
\end{equation}
is exactly satisfied for some $n$. Since the
sequence $\hat k_n$ is countable, and (\ref{2circl}),
in general, involves irrational requency ratios and
irrational coefficients, equation (\ref{2circl})
is only accidentially satisfied for some $n$ for a
measure zero set of graph parameters.
Hence, in general, even for marginal quantum graphs,
the points $\hat k_{n}$ still
serve as separating points and
the roots of the spectral equation can still
be obtained via the expansion
(\ref{kn}).

{\bf Example 7}. Star graph.
Another example of a marginal quantum graph is provided
by the star-graph shown in Fig.~9.
We consider the case with three different scaling potentials
on its three bonds and
Dirichlet boundary
conditions at the three dead ends. The spectral equation is
given by
\begin{eqnarray}
\cos(S_0k)=a_1\cos(S_1k)+a_2\cos(S_2k)+a_3\cos(S_3k)
\end{eqnarray}
where
\begin{eqnarray}
S_0&=S_{14}^0+S_{24}^0+S_{34}^0,\ \ \
S_{1}=S_{14}^0-S_{24}^0+S_{34}^0, \cr
S_{2}&=S_{14}^0-S_{24}^0-S_{34}^0,\ \ \
S_{3}=S_{14}^0+S_{24}^0-S_{34}^0
\end{eqnarray}
and
\begin{equation}
a_1={\beta_{14}-\beta_{24}+\beta_{34}
\over\beta_{14}+\beta_{24}+\beta_{34}},\ \ \
a_2={-\beta_{14}+\beta_{24}+\beta_{34}
       \over\beta_{14}+\beta_{24}+\beta_{34}},\ \ \
a_3={\beta_{14}+\beta_{24}-\beta_{34}
       \over\beta_{14}+\beta_{24}+\beta_{34}}.
\label{staras}
\end{equation}
It is straightforward to verify that
no matter what the sign of the individual $a_i$
in (\ref{staras}),
$\sum_{i=1}^{3}|a_{i}|=1$.
Therefore the condition (\ref{regmarg}) is satisfied
and the star-graph shown in Fig.~9 is another
example of a marginal quantum graph.
As discussed in the context of Example 6, and
with the exception of a set of measure zero of the
star-graph's
parameter space, the spectral expansion (\ref{kn}) is
still valid and can be used to obtain every single
one of the star-graph's eigenvalues individually, and
independently of all the other eigenvalues.

\section{Summary, Discussion and Conclusion}
Exact periodic orbit expansions for the global density of states
are known for many chaotic systems \cite{Roth,Selberg,AM}.
However, formula (\ref{kn}) is the first example of an
explicit expression for the {\em individual} quantum mechanical
levels, obtained {\em as a function of the level index} $n$,
for a classically chaotic system.
Additional explicit quantization formulas may be found
for other quantum graph systems, or even for quantum
systems unrelated to
quantum graphs as long as two
essential requirements are fulfilled. First
an exact periodic orbit
expansion for the density of states must
exist. Second it must be determined that
one of the system's
levels, $k_{*}$, is the only one in an interval
$\hat k_{*}'<k_{*}<\hat k_{*}''$.
Then one can always obtain the
corresponding periodic orbit expansion for $k_{*}$,
\begin{equation}
k_{*}=\int_{\hat k_{*}'}^{\hat k_{*}''}k\rho (k)dk,
\label{level}
\end{equation}
based on the periodic orbit expansion for $\rho(k)$.

It is reasonable to expect that generically there exist
separating points
$\hat k_{*}'$ and $\hat k_{*}''$ which separate every
$k_{*}$ from its neighbors, so that $k_{*}$ is the only root of
the spectral equation in the interval $[\hat k_{*}', \hat k_{*}'']$.
Hence expansions like (\ref{kn}) generally do exist.
However, knowing the positions of the separators $k'_{*}$ and
$k''_{*}$ around a particular level $k_{*}$, does not help
for finding
the separators for all the other levels.
So the most important
task for obtaining a general expression
for all the
levels of a quantum chaotic system is finding {\em a global
function for the separating points}
similar to (\ref{mid}), which
naturally enumerates the separators.
Therefore, even though it might be
possible to find, for some
systems, the separators for a {\em particular} quantum level
$k_{*}$, and then to obtain a
periodic orbit expansion for it according to
(\ref{level}), the expansion will work {\it only} for
level $k_{*}$
and will not represent a formula
which can be used for obtaining other levels.

The problem of finding a global
expression for the separating
points as a function of their
ordering index $n$ is directly related
to another well-known problem of
spectral theory of differential
operators, namely the problem of
approximating the staircase
function (\ref{N}) by a smooth average $\bar N(k)$.
Indeed, suppose there exists a
separating point $\hat k_{n}'$ between
every two roots of the spectral equation,
i.e. a solution of
the equation
\begin{equation}
\bar N(\hat k')=N(k),
\label{intersect}
\end{equation}
between every two roots of the spectral equation (similar to
(\ref{inverse}) and (\ref{cross})).
Since $\bar N(k)$ is a monotonic function,
the separating points can
then be found by inverting the
equation (\ref{intersect}),
\begin{equation}
\hat k_{N}=k(N),
\label{separate}
\end{equation}
in which the value of the staircase
function plays the role of the
separator index $\hat k_{N}$.
The equation (\ref{separate}) generalizes
(\ref{mid}), which can be used in (\ref{level}) to obtain
the periodic orbit expansions for all of the roots.

The smooth curve defined by
(\ref{separate}) where $N$ is considered to be
a continuous variable, intersects
every stair of the spectral staircase
(\ref{N}).
Unfortunately, finding a smooth function which approximates the
spectral staircase for a general differential operator with
generic boundary conditions is a rather complicated task.
It was proven by Weyl in 1912
that one can approximate $\bar N(k)$
by the phase-space volume of the system in question,
\begin{equation}
\bar N(E)\approx\int\Theta(E-H(x,p))
\frac{d^{D}xd^{D}p}{(2\pi\hbar)^{D}},
\label{Weyl}
\end{equation}
where $D$ is the dimensionality of the phase space, however this
average is certainly not guaranteed to satisfy the ``piercing
average'' condition (\ref{intersect}).
Since Weyl, this problem has been
addressed by numerous researchers (see, e.g.,
\cite{Hilf}), who succeeded in giving many improved estimates,
for $\bar N(k)$, but none of them a priori satisfy
(\ref{intersect}).

The important feature of the regular quantum graph systems is that
there exists a {\em global piercing average} (\ref{mid}), which
{\em uniformly} enumerates {\em all} the points separating one
root from another, and hence it is possible to obtain formula
(\ref{kn}) as a function of the index $n$.
In other words, the index $n$ in (\ref{kn}) is a quantum number,
and so the expression (\ref{kn}) for the energy levels of a chaotic
system in terms of classical periodic orbits can be considered as
a non-integrable analog of the EBK quantization scheme
\cite{Gutzw,Keller}.

It should be mentioned that despite the existence of a quantum
number $n$ in (\ref{kn}), the actual dependence of the energy
levels on the value of its quantum number is quite different from
the simple EBK scheme for integrable systems.
The expansion of the fluctuating part of the roots (\ref{average})
involves an intricate, conditionally convergent series and
is rather ``chaotic''.
The difference in complexity of the formulas (\ref{kn}) and
the EBK formula apparently reflects the complexity of the
geometry of the periodic orbits of the classically chaotic
quantum graphs.

Y.D. and R.B. gratefully acknowledge financial support by NSF grants
PHY-9900730 and PHY-9984075; Y.D. and R.V.J by NSF grant PHY-9900746.


\section{Appendix}
   For the sake of completeness, we present below a simple derivation
of the spectral determinant (\ref{det}), starting from the boundary
conditions at the vertex $V_{i}$:
\begin{equation}
\psi_{ij}(x)\mid_{x=0}=\varphi_{i}C_{ij}
\label{Aphase1}
\end{equation}
and
\begin{equation}
\sum_{j=1}^{N_{V}}C_{ij}\left(i\frac{d}{dx_{ij}}+A_{ij}\right)
\psi_{ij}(x_{ij})\mid_{x=0}=\lambda_{i}\varphi_{i}.
\label{Aphase2}
\end{equation}
We would like to present the wave function,
\begin{equation}
\psi_{ij}(x)=\frac{1}{\sqrt{\beta_{ij}k}}
\left(a_{ij}e^{-i\beta_{ij}kx}+b_{ij}e^{i\beta_{ij}kx}\right) ,
\label{Apsi}
\end{equation}
which satisfies these boundary conditions as a superposition of the
partial waves
\begin{equation}
\psi^{(i)}_{jj'}(x_{j})=
\delta_{jj'}\frac{e^{i\left(-\beta_{ij}k+A_{ij}\right)x_{j}}}
{\sqrt{\beta_{ij}k}}
+
\sigma_{ji,ij'}\frac{e^{i\left(\beta_{ij}k+A_{ij}\right)
x_{j}}}{\sqrt{\beta_{ij}k}}.
\label{scatt}
\end{equation}
scattering on the vertices of the graph.
So
\begin{equation}
\psi_{ij}(x_{j})=\sum_{j'=1}^{N_{V}}a_{ij'}\psi_{j,j'}^{(i)}(x_{j})=
\frac{a_{ij}}{\sqrt{\beta_{ij}k}}e^{-i\left(\beta_{ij}k-A_{i,j}\right)x_{j}}+
\frac{e^{i\left(\beta_{ij}k+A_{i,j}\right) x_{j}}}{\sqrt{\beta_{ij}k}}
\sum_{j'=1}^{N_{V}}a_{ij'}\sigma_{ji,ij'},
\label{Apssi}
\end{equation}
with the appropriate weights $a_{ij'}$ corresponding to the incoming
flux on the bond $B_{j'i}$ towards the vertex $V_{i}$.
Comparing this expression to (\ref{Apsi}) yields
\begin{equation}
b_{ij}=\sum_{j'=1}^{N_{V}}\sigma_{ji,ij'}a_{ij'}.
\label{ab1}
\end{equation}
Substituting (\ref{Apssi}) into the boundary conditions (\ref{Aphase1})
and (\ref{Aphase2}) at the vertex $V_{i}$ we get correspondingly
\begin{equation}
\sum_{j'=1}^{N_{V}}\frac{a_{ij'}}{\sqrt{\beta_{ij}k}}\left(\delta_{jj'}+
\sigma_{ji,ij'}\right) =\varphi_{i}C_{ij}
\label{Acont}
\end{equation}
and
\begin{equation}
\sum_{j,j'=1}^{N_{V}}C_{ij}a_{ij'}\sqrt{\beta_{ij}k}
\left(\delta_{jj'}-\sigma_{ji,ij'}\right)
=i\lambda_{i}\varphi_{i},
\label{Aflux}
\end{equation}
Substituting (\ref{Acont}) into (\ref{Aflux}) one gets
\begin{equation}
C_{ij}\sum_{l,j'=1}^{N_{V}}C_{il}a_{ij'}\sqrt{\beta_{il}k}
\left(\delta_{lj'}-\sigma_{l,j'}^{(i)}\right) =i\lambda
_{i}\sum_{j'=1}^{N_{V}}\frac{a_{ij'}}{\sqrt{\beta_{ij}k}}
\left(\delta_{jj'}+\sigma_{ji,ij'}\right)
\label{Ascatte}
\end{equation}
In case of the linear scaling, $\lambda_{i}=k\,\lambda_{i}^{0}$, this
yields
\begin{equation}
\sum_{j'=1}^{N_{V}}a_{ij'}C_{ij}\sum_{l=1}^{N_{V}}C_{il}
\sqrt{\beta_{il}}\left(
\delta_{lj'}-\sigma_{li,ij'}\right) =
i\lambda_{i}^{0}\sum_{j'=1}^{N_{V}}\frac{a_{ij'}}
{\sqrt{\beta_{ij}}} \left(\delta_{jj'}+\sigma_{ji,ij'}\right)
\end{equation}
Comparing the coefficients in front of $a_{ij'}$, we get
\begin{equation}
C_{ij}\sum_{l=1}^{N_{V}}C_{il}\delta_{lj'}\sqrt{\beta_{il}}
-C_{ij}\sum_{l=1}^{N_{V}}C_{il}\sqrt{\beta_{il}}\sigma_{l,j'}^{(i)}
-i\lambda_{i}^{0}\frac{\delta_{jj'}}{\sqrt{\beta_{ij}}}=
\frac{i\lambda_{i}^{0}}{\sqrt{\beta_{ij}}}\sigma_{ji,ij'}
\end{equation}
or, after performing the summation over $l$,
\begin{equation}
C_{ij}C_{ij'}\sqrt{\beta_{ij'}}-C_{ij}\Gamma_{i,j'}^{i}
-i\lambda_{i}^{0}\frac{\delta_{jj'}}
{\sqrt{\beta_{ij}}}= \frac{i\lambda_{i}^{0}}{\sqrt{\beta_{ij}}}
\sigma_{ji,ij'}  \label{ur}
\end{equation}
where $\Gamma_{i,j'}^{i}=\sum_{l=1}^{N_{V}}C_{il}\sqrt{\beta_{il}}
\sigma_{li,ij'}$.

Multiplying both sides by $C_{ij}\beta_{ij}$ and summing over $j$
yields
\begin{equation}
v_{i}C_{ij'}\sqrt{\beta_{ij'}}-v_{i}\Gamma_{i,j'}^{i}
-i\lambda_{i}^{0}C_{ij'}\sqrt{\beta_{ij'}}
=i\lambda_{i}^{0}\Gamma_{i,j'}^{i},
\end{equation}
where $v_{i}=\sum_{j}C_{ij}\beta_{ij}$. Hence
\begin{equation}
\frac{v_{i}-i\lambda_{i}^{0}}{v_{i}+i\lambda_{i}^{0}}C_{ij'}
\sqrt{\beta_{ij'}}=\Gamma_{i,j'}^{i},
\end{equation}
which can be used in (\ref{ur}) to obtain
\begin{equation}
C_{ij}C_{ij'}\sqrt{\beta_{ij'}}-C_{ij}\frac{v_{i}-
i\lambda_{i}^{0}}{v_{i}+i\lambda_{i}^{0}}C_{ij'}
\sqrt{\beta_{ij'}}- i\lambda_{i}^{0}\frac{\delta_{jj'}}
{\sqrt{\beta_{ij}}}= \frac{i\lambda_{i}^{0}}{\sqrt{\beta_{ij}}}
\sigma_{ji,ij'}
\end{equation}
or
\begin{equation}
\sigma_{ji,ij'}=\left(-\delta_{jj'}+
\frac{2\sqrt{\beta_{ij}\beta_{ij'}}}{v_{i}+i\lambda_{i}^{0}}\right)
C_{ji}C_{ij'}.
\label{sigma}
\end{equation}
We see that in the scaling case the matrix elements $\sigma_{ji,ij'}$
of the vertex scattering matrix $\sigma$ are $k$-independent
constants.

The matrix element $\sigma_{ji,ij}$ has the meaning of the reflection
coefficient from the vertex $V_{i}$ along the bond $B_{ij}$, and the
elements $\sigma_{ji,ij'}$, $j\neq j'$ are the
transmission coefficients for transitions between different bonds.
The equation (\ref{ab1}) can be written as
\begin{equation}
\vec{b}=\tilde{T}\vec{a},
\label{Tmat}
\end{equation}
where
\begin{equation}
\tilde{T}\equiv \tilde{T}_{ij,nm}=\delta_{in}C_{ji}C_{nm}\sigma_{ji,im}.
\label{tilt}
\end{equation}

In the symmetrical basis $\psi_{ji}\left(L_{ij}-x\right) =\psi_{ij}(x)$
one has
\begin{equation}
\psi_{ji}\left(L_{ij}-x\right) =
a_{ji}\frac{e^{i\left(-\beta_{ij}k+A_{ji}\right)
\left(L_{ij}-x\right)}}{\sqrt{\beta_{ij}k}}+
b_{ji}\frac{e^{i\left(\beta_{ij}k+A_{ji}\right)
\left(L_{ij}-x\right)}}{\sqrt{\beta_{ij}k}}=\psi_{ij}(x),
\end{equation}
so the coefficients $a_{ij}$ and $b_{ij}$ are related as
\begin{equation}
a_{ji}=b_{ij}e^{i\left(\beta_{ij}k+A_{ij}\right) L_{ij}},\ \ \
b_{ji}=a_{ij}e^{i\left(-\beta_{ij}k+A_{ij}\right) L_{ij}}.
\label{ab}
\end{equation}
The coefficients $a_{ij}$ and $a_{ji}$, ($b_{ij}$ and $b_{ji}$) are
considered different, so the bonds of the graph are ``directed''.

Equations (\ref{ab}) can be written in matrix form,
\begin{equation}
\vec{a}=P\tilde{D}(k)\vec{b},
\label{Dmat}
\end{equation}
where $\vec{a}$ and $\vec{b}$ are the $2N_{B}$ dimensional vectors
of coefficients and $\tilde{D}$ is a diagonal matrix in the
$2N_{B}\times 2N_{B}$ space of directed bonds,
\begin{equation}
\tilde{D}_{ij,pq}(k)=\delta_{ip}\delta_{jq}
e^{i\left(\beta_{ij}k+A_{ij}\right)L_{ij}},
\label{tild}
\end{equation}
and
\begin{equation}
P=\pmatrix{0 & 1_{N_{B}} \cr 1_{N_{B}} & 0},
\label{P}
\end{equation}
where $1_{N_{B}}$ is the $N_{B}$-dimensional unit matrix. The pairs of
indices $(ij)$, $(pq)$, identifying the bonds of the graph $\Gamma $, play
the role of the indices of the matrix $\tilde{D}(k)$.

Equations (\ref{Dmat}) and (\ref{Tmat}) together result in
\begin{equation}
\vec{a}=S(k)\vec{a},
\label{lin}
\end{equation}
with the matrix $S(k)$ (the total graph scattering matrix) given by
\begin{equation}
S(k)=D(k)T,
\label{S}
\end{equation}
where $D=P\tilde{D}P$ and $T=P\tilde{T}$.

\pagebreak

\centerline{\bf Figure Captions}

\bigskip\noindent
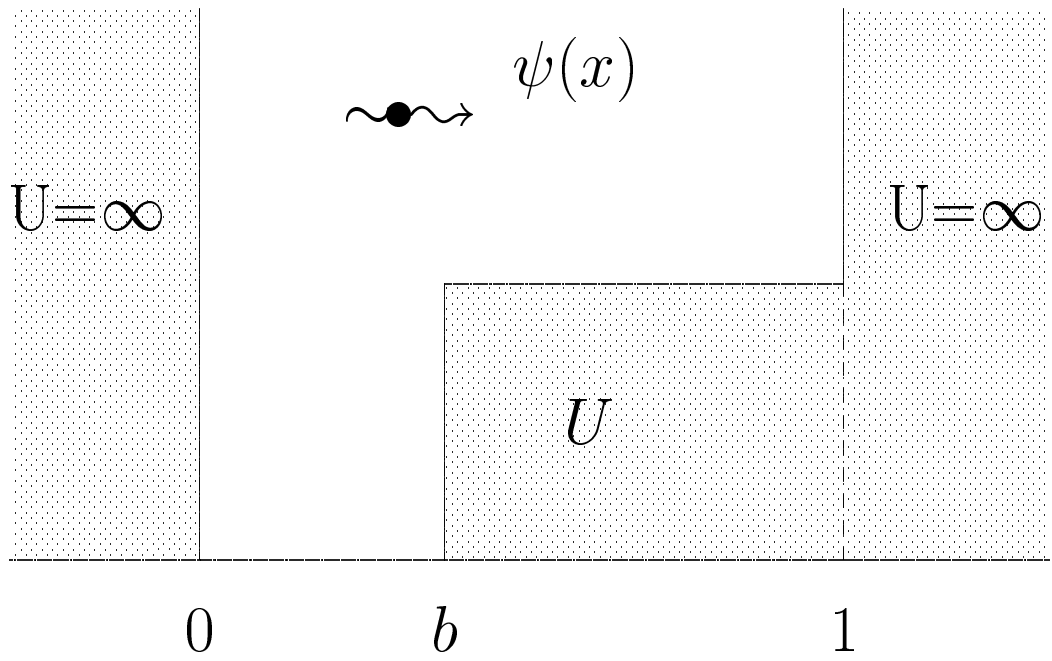
{\bf Fig.~1:} Sketch of a step-potential in a box, a well-known
textbook quantum problem.

\bigskip\noindent
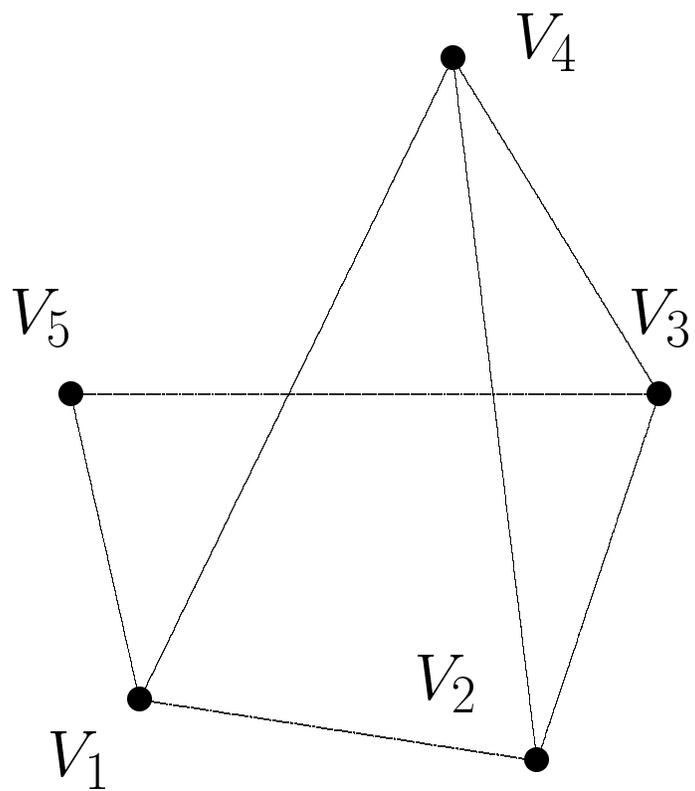
{\bf Fig.~2:} Sample graph with five vertices and seven bonds.

\bigskip \noindent
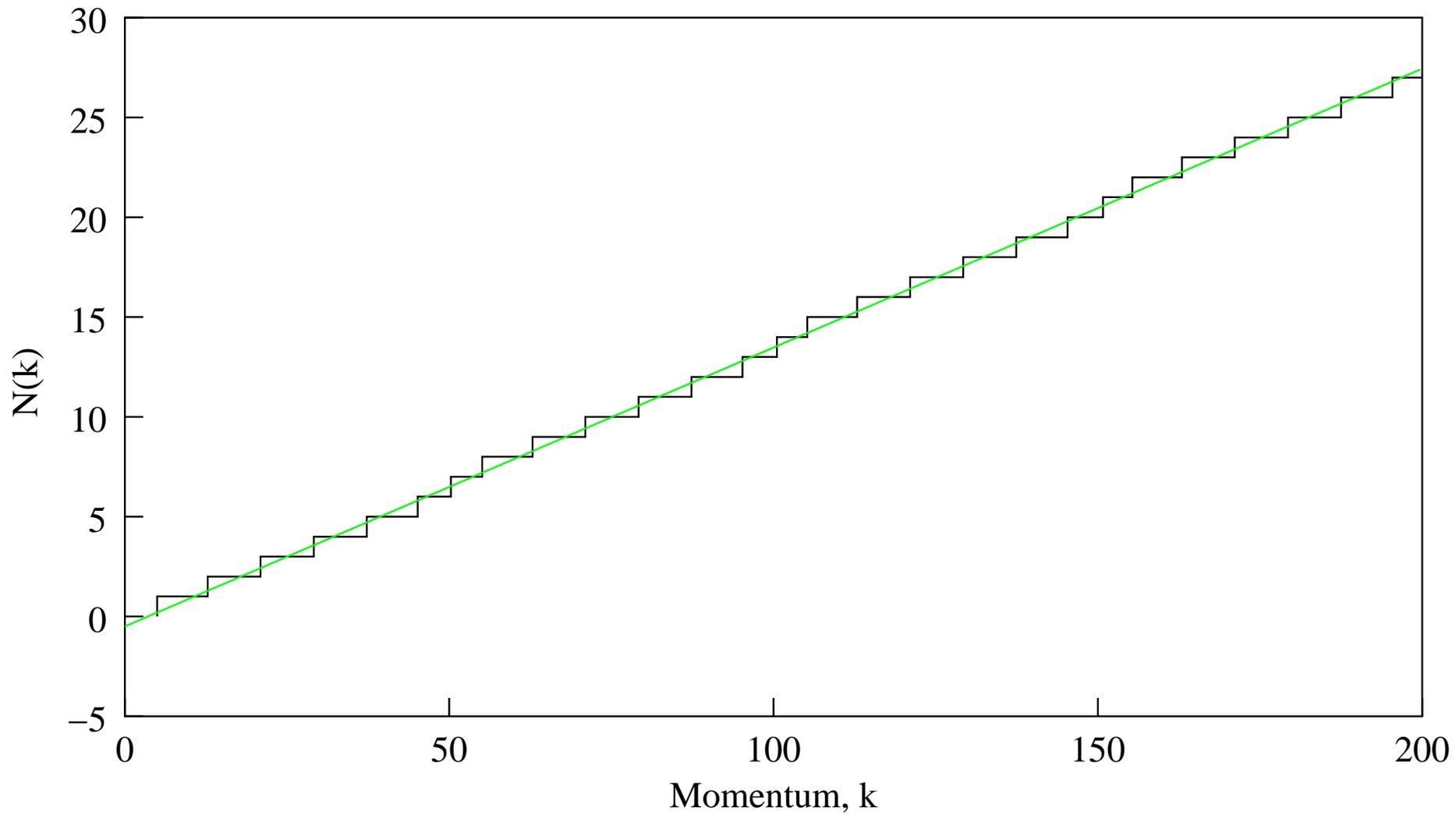
{\bf Fig.~3:} The exact
spectral staircase $N(k)$ and its
average $\bar N(k)$ for the scaling step-potential
shown in Fig.~1 with $b=0.3$ and $\lambda=1/2$.
The average $\bar N(k)$
crosses every ``stair'' of $N(k)$
(piercing average) at the equally
spaced separating points $\hat k_{n}$.

\bigskip \noindent
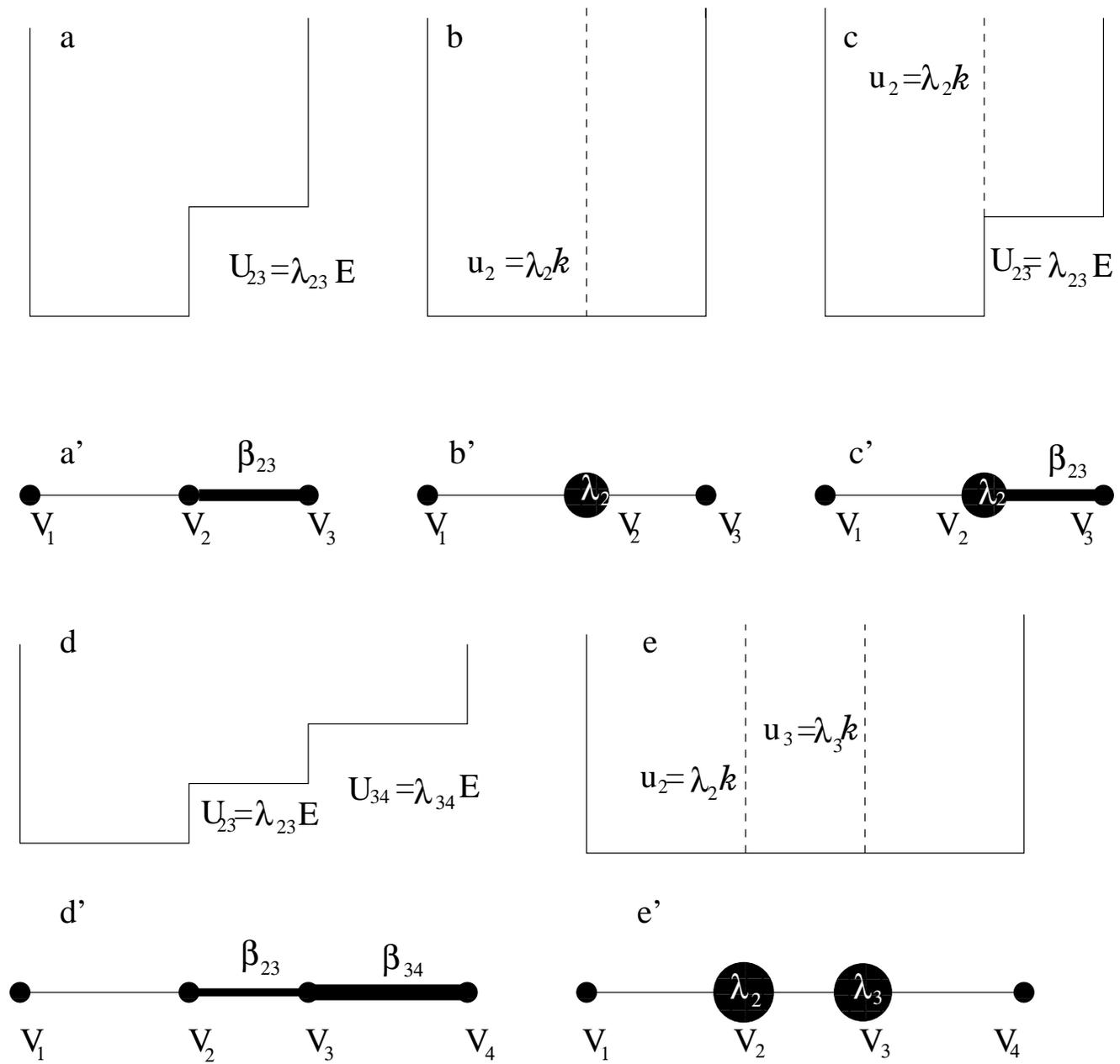
{\bf Fig.~4:} Collection of potentials and their
associated linear quantum graphs that
serve as examples to illustrate
the concept of regular quantum graphs.
(a) Scaling step-potential
in a box and its associated
three-vertex linear graph (a'). (b)
Scaling $\delta$-function in a box and its corresponding
three-vertex linear graph (b').
Combined scaling $\delta$-function
and step potential in a box (c) with its linear
three-vertex quantum
graph (c'). Two scaling steps (d)
and two scaling $\delta$-functions
(e) in a box together with their associated
four-vertex dressed
linear quantum graphs (d') and (e'), respectively.

\bigskip \noindent
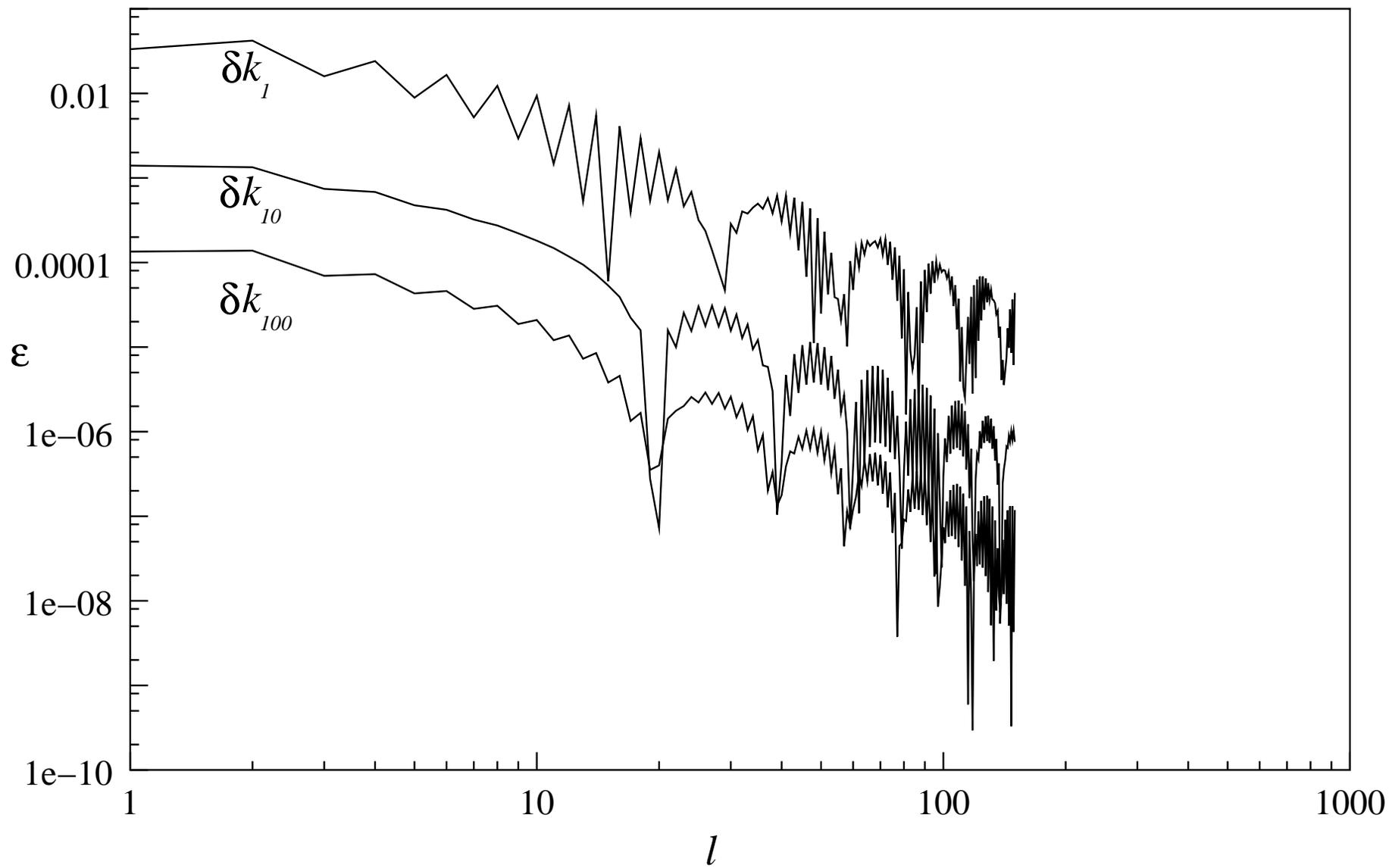
{\bf Fig.~5:} Comparison between the exact eigenvalues
$k_n$ and the $k_n$-values computed via (\ref{kn})
for the scaling step-potential shown in Fig.~1.
Shown is the
relative error
$\epsilon_n^{(l)}=|k_n^{(l)}-k_n|/k_n$,
$n=1,10,100$,
of the result $k_n^{(l)}$ predicted
by (\ref{kn}) compared to the numerically
obtained exact result $k_n$ as a
function of the binary code length $l$ of
the orbits used in the expansion (\ref{kn}).
We used $b=0.3$, $\lambda=1/2$.

\bigskip \noindent
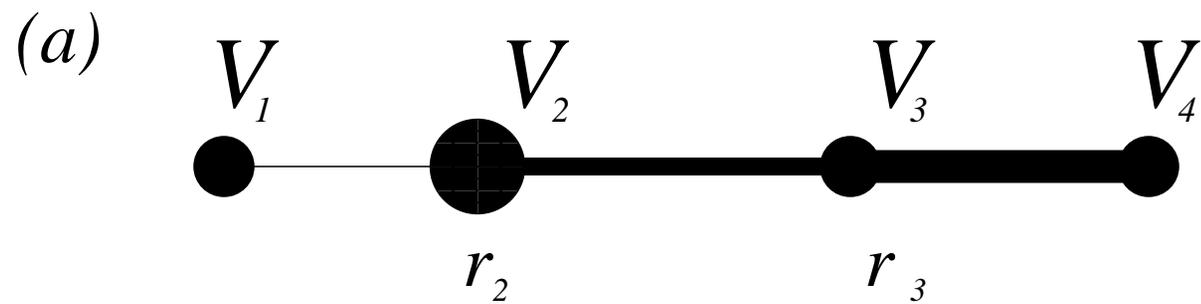
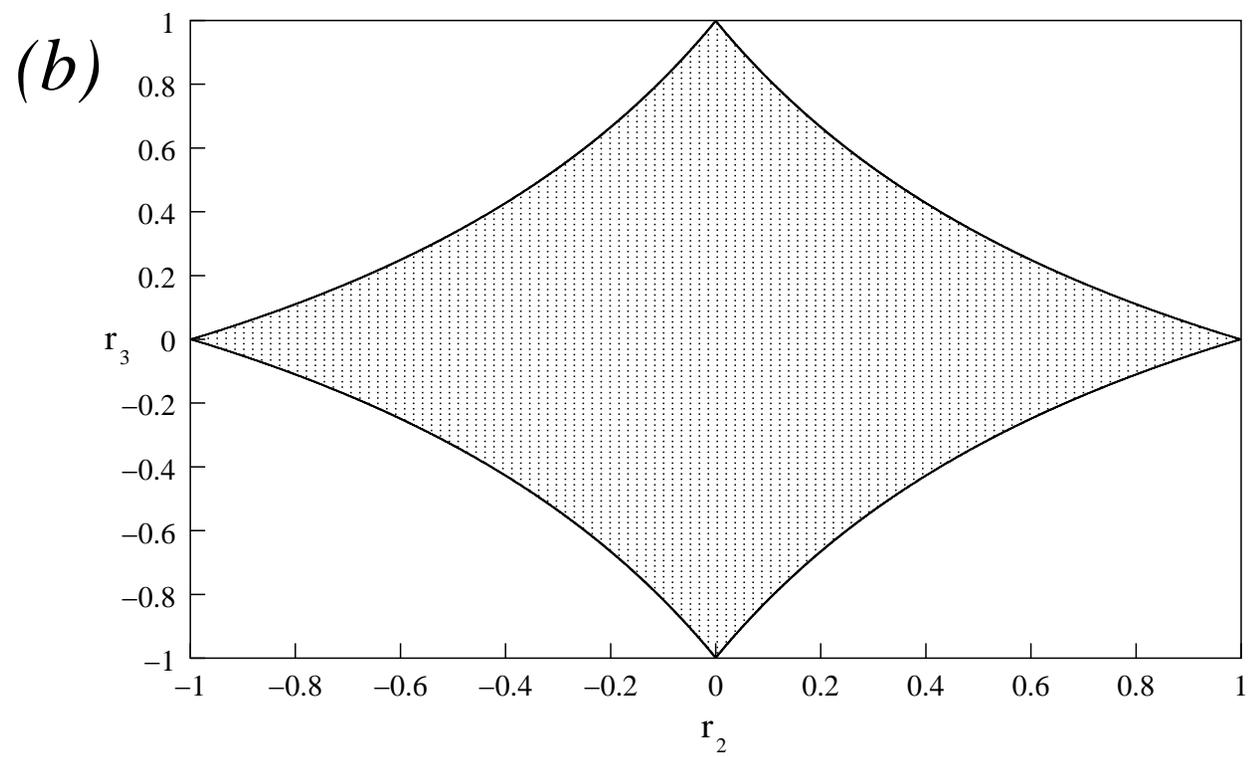
{\bf Fig.~6:} Four-vertex linear chain graph (a) and the
corresponding space $(r_2,r_3)$ of reflection
coefficients (b).
The shaded region in the $(r_2,r_3)$ space
corresponds to the regular regime of the
quantum graph shown in (a).
This demonstrates that the subset of
regular quantum graphs within the set
of all four-vertex linear quantum graphs
is non-empty and of finite measure.

\bigskip \noindent
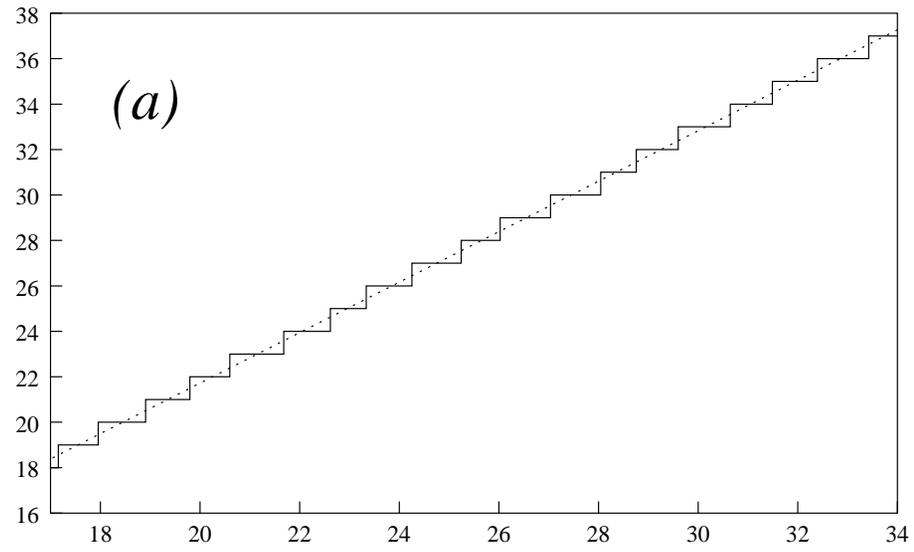
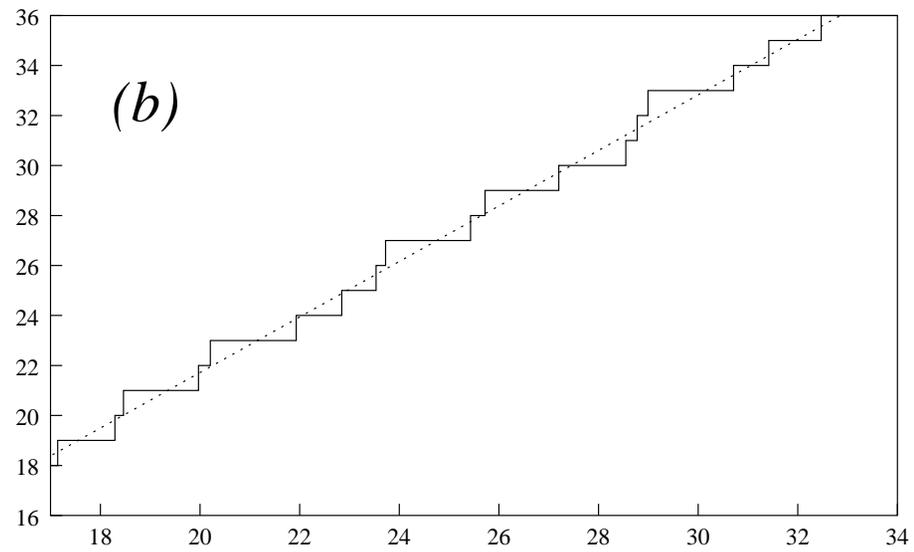
{\bf Fig.~7:} The exact spectral
staircase $N(k)$ and its
average $\bar N(k)$ for the
regular $r_{2}=0.2$, $r_{3}=0.3$) (a)
and the irregular $r_{2}=0.98$,
$r_{3}=0.99$) (b) regimes of the
four-vertex linear graph shown in Fig.~6a.
In the regular regime (a) the average
staircase $\bar N(k)$ pierces every step of
$N(k)$. This is not the case in (b), characteristic
of the irregular regime.

\bigskip \noindent
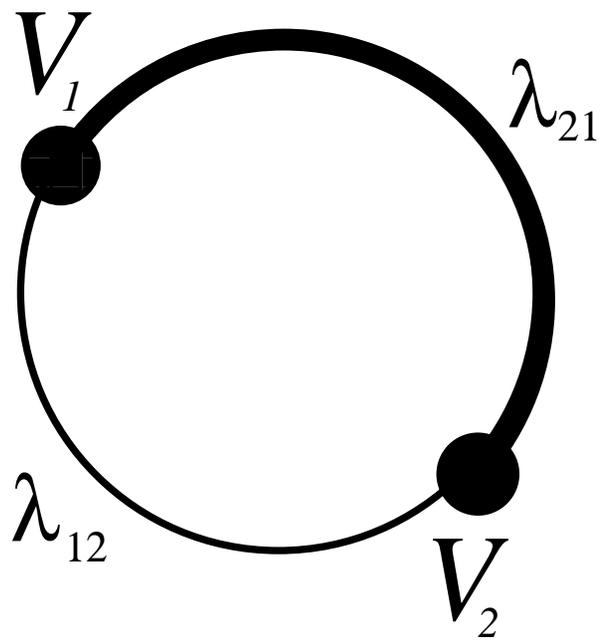
{\bf Fig.~8:} Two-vertex circular graph.
              In the mathematically positive sense,
              $\beta_{12}$ is the scaling coefficient
              of the bond connecting vertex $V_1$ with
              vertex $V_2$, $\beta_{21}$ is the
              scaling coefficient of the bond connecting
              $V_2$ with $V_1$.
              This labelling is possible only in the absence
              of a magnetic field ($A_{ij}=0$) where
              the sense of traversal of a bond is irrelevant.

\bigskip \noindent
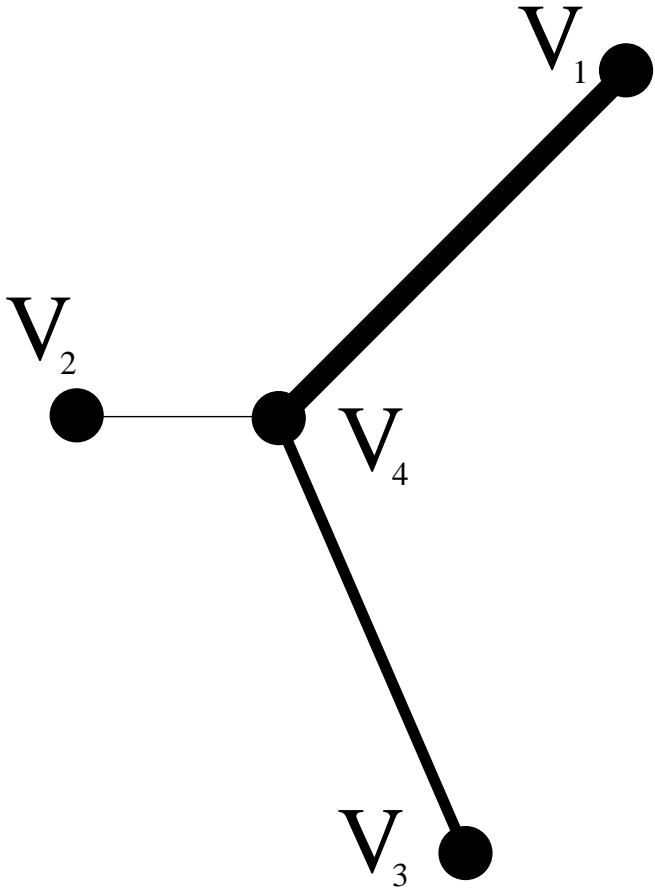
{\bf Fig.~9:} Scaling star-graph with three bonds and
              four vertices.

\bigskip \noindent
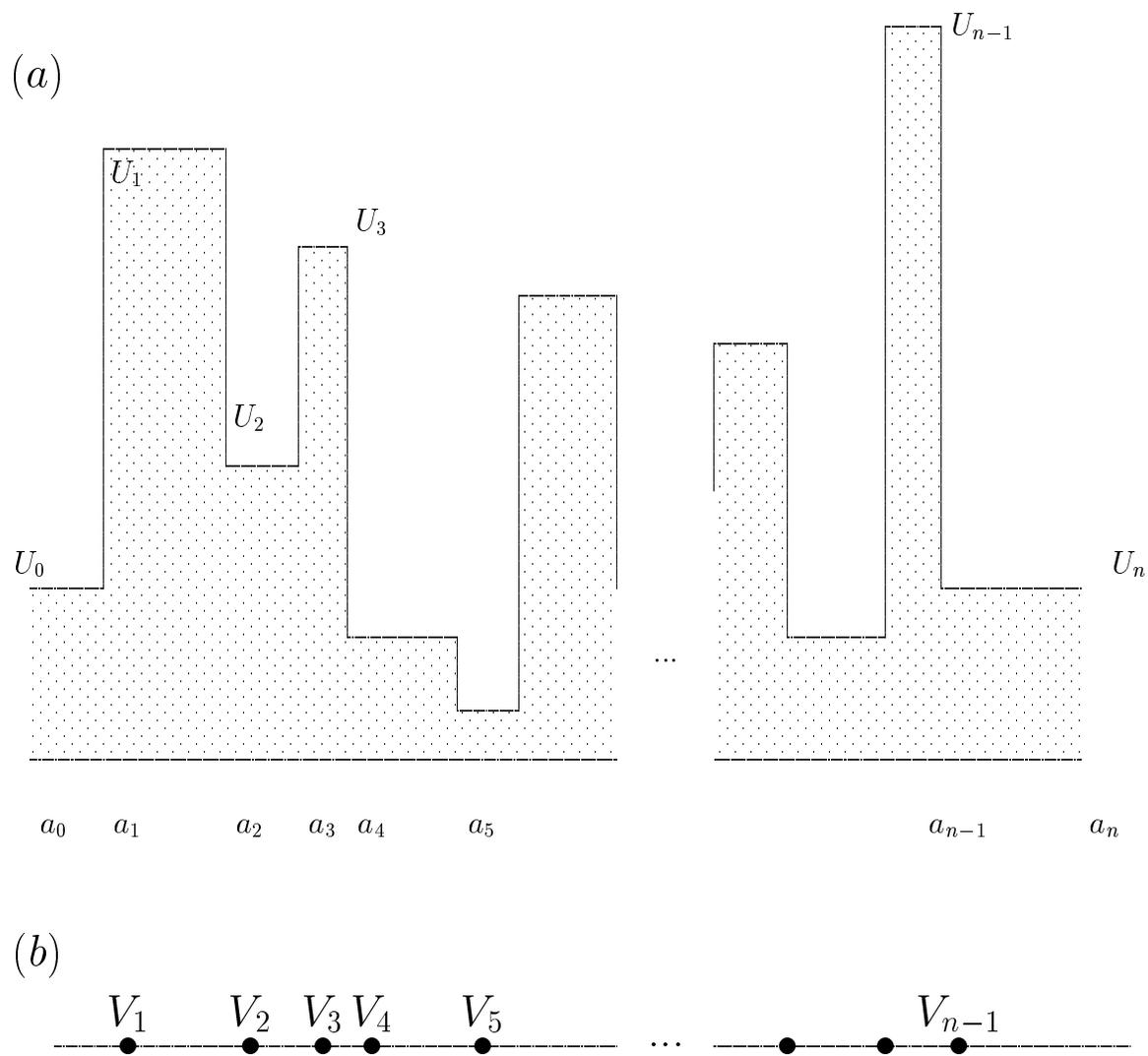
{\bf Fig.~10:} Sketch of a piecewise constant potential
               (``Manhattan potential'') (a) and its
               associated linear graph (b).

\end{document}